\documentstyle[preprint,epsfig,aps]{revtex}
\input epsf.tex \def\DESepsf(#1 width #2){\epsfxsize=#2 \epsfbox{#1}}
\begin{document}
%
\draft 
\title{CP-odd $A^0$ production at $e^+e^-$ colliders in MSSM with 
CP violating phases}
\author{Abdesslam Arhrib\footnote{e-mail arhrib@mppmu.mpg.de}} 
\address{ Max-Planck Institut f\"ur Physik, F\"ohringer Ring 6, 
80805 M\"unchen, Germany\\
and\\
LPHEA, Physics Department, Faculty of Science-Semlalia, P.O.Box 2390, \\
Marrakesh, Morocco.}

\tighten

\date{\today} 
\maketitle
\newcommand{\poss}[2] { \left\{ \!\! 
   {\footnotesize\begin{array}{r} #1 \\ #2 \end{array}} \!\!\right\}}
\def\ga{\mathrel{\raise.3ex\hbox{$>$\kern-.75em\lower1ex\hbox{$\sim$}}}}
\def\la{\mathrel{\raise.3ex\hbox{$<$\kern-.75em\lower1ex\hbox{$\sim$}}}}


\newcommand{\nn}{\noindent}
\renewcommand{\thefootnote}{\fnsymbol{footnote}}


\begin{abstract} 
We study the production of a heavy CP-odd $A^0$ boson in association with
a photon $e^+e^-\to A^0\gamma$ and a Z boson $e^+e^-\to  A^0 Z$
as well as the single production of $A^0$ via $e^+e^- \to \nu_e 
\bar{\nu}_e A^0$
in the MSSM with CP violating phases. 
In the case of $e^+e^-\to A^0\gamma / A^0 Z$,
we show that the squark contribution,
which vanishes in the MSSM with real parameters, 
turns out to be sizeable in presence of CP violating 
phases in the soft SUSY parameters.
For $e^+e^- \to \nu_e \bar{\nu}_e A^0$ in both the 2HDM and MSSM with real
parameters, the cross section does not reach observable rates at a NLC.
It is found that with a large CP violating phase for $A_t$, 
cross sections of the order 0.08 fb are attainable for all the processes 
$e^+e^-\to A^0\gamma$, $e^+e^-\to  A^0 Z$ and 
$e^+e^- \to \nu_e \bar{\nu}_e A^0$.
\end{abstract}

\vfill

\newpage

\pagestyle{plain}
\renewcommand{\thefootnote}{\arabic{footnote} }
\setcounter{footnote}{0}

\section*{1.~Introduction}

Supersymmetric (SUSY) theories, in particular the Minimal Supersymmetric
Standard Model (MSSM) \cite{MSSM}, are currently considered as the 
most theoretically well motivated extensions of the Standard Model. 
Recently, the phenomenology of the MSSM with complex SUSY parameters
has received growing attention \cite{MSSMCP0,MSSMCP2,MSSMCP1}.
Such phases give new sources of CP violation 
which may explain: electroweak baryogenesis scenarios \cite{Baryogen}, 
and CP violating phenomena in K and B decays \cite{baek}.
It has been shown in \cite{dugan} 
that by assuming universality of the gaugino masses at
a high energy scale, the effects of complex soft SUSY 
parameters in the MSSM can be parametrized 
by two independent CP violating phases: the phase of the 
Higgsino mass term $\mu$ (Arg($\mu$)) and the phase 
of the trilinear scalar coupling parameters $A=A_f$ (Arg($A_f$)) 
of the sfermions $\widetilde{f}$. 
The presence of large SUSY phases can give contributions to 
electric dipole moments of the electron and neutron (EDM) 
which exceed the experimental upper bounds. In a variety of SUSY 
models such phases turn out to be severely constrained 
by such constraints i.e. ${\rm Arg}(\mu) < {\cal }(10^{-2})$ for a SUSY
mass scale of the order of few hundred GeV \cite{nath,savoy}.

However, the possibility of having large CP violating phases can still be consistent with experimental data in any of the following three scenarios: i) Effective SUSY models \cite{nath},
ii) Cancellation mechanism \cite{cancell} and iii) Non-universality of 
trilinear couplings $A_f$ \cite{trilin}.
At tree level, the Higgs sector of the MSSM with complex phases is
CP conserving. The particle spectrum consists of 
2 neutral CP--even scalars ($h^0$ and $H^0$),
a pair of charged scalars  ($H^+,H^-$) and 
a CP--odd neutral scalar $A^0$. At the one loop level the Higgs 
sector of the MSSM is no longer CP conserving.
The presence of SUSY phases induces mixing between the CP--even and 
CP--odd scalars, resulting in the 3 mass eigenstates $H_1^0$,
$H_2^0$ and $H^0_3$ which do not have a definite CP parity.
This mixing affects their phenomenology at present
and future colliders, both in production mechanisms 
and decay partial widths \cite{BR0,BR}. Recent studies \cite{AA,demir} 
show that the processes $e^+e^-\to Z^*\to H^0_iZ$ and 
$e^+e^-\to H^0_i\nu_e\overline \nu_e$
would be a way of probing CP violation in the Higgs sector at the NLC.
Both mechanisms are mediated by the effective tree--level
couplings $H^0_iVV$, $V=Z,W^\pm$.\\
Due to electromagnetic invariance, there is no effective tree level
coupling $\gamma Z H_i^0$ and $\gamma \gamma H_i^0$. 
Consequently, the process $e^+e^- \to \gamma H_i^0$ is loop 
mediated. To have an idea about the order of magnitude of 
the cross section of $e^+e^- \to \gamma H_i^0$, we will evaluate
the cross section of $e^+e^- \to \gamma A^0$ in MSSM including 
CP violating phases of $A_{t,b}$.
Such approximation would be correct only in the case where the 
couplings $H_i^0f\bar{f}\approx A^0f\bar{f}$, 
which means that CP violating phases does not affect that 
much the $H_i^0f\bar{f}$ couplings. Same arguments apply for 
$e^+e^- \to Z H_i^0$ and $e^+e^- \to \nu_e \overline{\nu}_e H_i^0$
specially when $H_i^0$ is weakly interacting with gauge bosons
$H_i^0 VV\approx A^0VV=0$.

In the CP conserving MSSM 
we have access only to $e^+e^-\to Z^*\to h^0Z/H^0Z$ and 
$e^+e^-\to \nu_e\overline \nu_e h^0/\nu_e\overline \nu_e H^0$ 
while $e^+e^-\to A^0Z$ and $e^+e^-\to \nu_e\overline \nu_e A^0$ 
are mediated at one loop level. 
The one loop evaluation of $e^+e^-\to A^0Z$ has been studied
both in the MSSM with real parameters \cite{AAC1} and 
the 2HDM \cite{AAC2,LG}.
Note that ref.\cite{LG} has also studied the heavy fermion contribution 
to $e^+e^-\to \nu_e\overline \nu_e A^0$ through W--W fusion.
In order to be able to disentangle the CP conserving MSSM 
from the CP violating one, the evaluation of the cross 
section of $e^+e^-\to A^0Z$ and 
$e^+e^-\to \nu_e\overline \nu_e A^0$ is necessary
and this is the goal of the present paper. 
Such processes, if observed, can give information on the
vertices $A^0\gamma\gamma$, $A^0\gamma Z$, $A^0ZZ$ and $A^0WW$.

CP-odd Higgs bosons can be produced at $e^+e^-$ colliders 
\cite{eecolliders} via $e^+e^- \to h^0 A^0\ , \ H^0 A^0$ and 
$e^+e^- \to b\overline bA^0,t\overline tA^0$ \cite{Dawson}.
If the $\gamma\gamma$ option is available then $\gamma\gamma\to A^0$
is very promising for $\tan\beta<10$ \cite{asner}.
In $\gamma\gamma$ colliders, one can have access to CP-odd $A^0$
through the associate production $\gamma\gamma \to Z A^0$
or the pair production $\gamma\gamma \to A^0 A^0$ \cite{gounaris}.
Note that in the MSSM the kinematically favored mechanism 
$e^+e^-\to Z^*\to A^0h^0$ is suppressed by the factor
$\cos^2(\beta-\alpha)$ which is very small in the 
region $M_A\ge 250$ GeV, while 
$e^+e^-\to Z^*\to A^0H^0$ is kinematically suppressed.

At future $e^+e^-$ colliders the simplest way to produce 
CP--even Higgs scalars is in the Higgsstrahlung process 
$e^+e^-\to Z^*\to H Z$. Due to CP-invariance, 
the CP--odd $A^0$ possesses no
tree-level coupling $A^0$-Z-Z and $A^0$-Z-photon. The
dominant contribution is therefore from higher order diagrams
which will be mediated by both SM and non--SM particles.
Therefore the rates are expected to be strongly model dependent.
Note that the associate production of single Higgs 
boson  with gauge boson: $H^\mp$-$W^\pm$ \cite{WH}, 
$A^0$-$Z$, $A^0$-$\gamma$ and $\gamma H^0$ 
\cite{abdel,AAC1,AAC2,LG}
would allow greater kinematical reach for the 
Higgs bosons mass (up to $\sqrt{s}-M_V$, $M_V$ is the mass
of the gauge boson).
If any of $e^+e^-\to A^0\gamma $, $e^+e^- \to A^0Z$ 
and/or $e^+e^-\to \nu_e\overline \nu_e A^0$
were sizeable it would provide an alternative 
way of producing $A^0$ at $e^+e^-$ colliders,
with a kinematical reach superior to that
for the mechanism $e^+e^-\to Z^*\to A^0H^0$.

In this paper, we calculate the associated production
mechanisms $e^+e^-\to A^0 \gamma$ and $e^+e^-\to A^0Z$ 
in the MSSM, numerical results are given both for 
polarized and unpolarized electron-positron beams.
We analyze also the one-loop process $e^+e^-\to 
\nu_e\overline \nu_e A^0$ in 2HDM and MSSM.
Heavy fermions contribution to $e^+e^-\to 
\nu_e\overline \nu_e A^0$ one-loop W boson fusion production 
has been evaluated in \cite{LG}. We complete this study by 
including fermions, sfermions, 
charginos and neutralinos contributions in different topologies 
and not only for W boson fusion. 
The effect of CP phases of trilinear Soft SUSY 
breaking parameters $A_{t,b}$ on $e^+e^- \to \gamma A^0$, 
$e^+e^- \to Z A^0$  and $e^+e^- \to \nu_e\overline \nu_e A^0$
is examined. We provide also comparison of 
$e^+e^- \to Z A^0 \ , \nu_e\overline \nu_e A^0$ in CP conserving 
MSSM with $e^+e^- \to Z H_i^0 \ , \nu_e\overline \nu_e H_i^0$ in 
CP violating MSSM \cite{MSSMCP0,MSSMCP2,MSSMCP1}.

\section*{2.~Notation and conventions}
First we summarize the MSSM parameters needed in our analysis,
with particular attention given to the sfermion sector. 
In the MSSM, the sfermion sector is specified by the mass matrix in 
the basis $(\tilde{f}_L^{},\tilde{f}_R^{})$. In terms of 
the scalar mass $\widetilde{M}_L$, $\widetilde{M}_R$, 
the Higgs-Higgsino mass parameter $\mu$ and
the soft SUSY-breaking trilinear coupling $A_f$, the sfermion mass 
matrices squared reads as \cite{gun}: 
\begin{equation}
{\cal M}^2_{\tilde{f}}= 
     \left( \begin{array}{cc} 
                m_f^2 + m_{LL}^2 & m_{LR}^* m_f \\
                m_{LR} m_f     & m_f^2 + m_{RR}^2
            \end{array} \right) \label{eq1}
\end{equation}
with
\begin{eqnarray}
  m_{LL}^2 &=& \widetilde{M}_{L}^2 
    + m_Z^2\cos 2\beta\,( I_3^f - Q_f s_W^2 ) ,     \label{eq:b} \\
  m_{RR}^2 &=& \widetilde{M}_{R}^2  
                  + m_Z^2 \cos 2\beta\, Q_f s_W^2 , \label{eq:c} \\[2mm]
  m_{LR}    &=& 
  (A_f - \mu^* (\tan\beta)^{-2I_3^f}) \,\, . \label{eq:d}
\end{eqnarray}
$I_3^f=\pm 1/2$ and $Q_f$ are the weak isospin and electric charge of 
the sfermion $\tilde{f}$ and $\tan\beta = \frac{v_2}{v_1}$ with 
$v_1$, $v_2$ being the vacuum expectation value of the Higgs fields. 

We will take $A_f$ and $\mu$ as complex parameters: 
$A_f = |A_f| \, e^{i{\rm arg}({A_f})}$ 
and $\mu = |\mu| \, e^{i{\rm arg}({\mu})}$ with 
$0 < {\rm arg}{(A_f,\mu)} \leq \pi $. 
The hermitian matrix (\ref{eq1}) is then diagonalised by 
a unitarity matrix 
$R_{{\tilde{f}}}$, which rotates the current eigenstates,
${\tilde{f}}_L$ and ${\tilde{f}}_R$, into the mass
eigenstates $\tilde{f}_1$ and $\tilde{f}_2$ as follows:
\begin{equation}
\left( \begin{array}{c} 
                {\tilde{f}}_1 \\
                {\tilde{f}}_2
\end{array} \right) = R^{{\tilde{f}}} 
\left( \begin{array}{c} 
{\tilde{f}}_L \\
{\tilde{f}}_R
\end{array} \right) = 
\left( \begin{array}{cc} 
e^{i\delta_{f}/2} \, \cos{\theta_f}  & 
e^{-i\delta_{f}/2}\, \sin{\theta_f}   \\
-e^{i\delta_{f}/2}\, \sin\theta_f &
e^{-i\delta_{f}/2}\, \cos\theta_f
            \end{array} \right) 
\left( \begin{array}{c} 
                {\tilde{f}}_L \\
                {\tilde{f}}_R
\end{array} \right) \label{eqe}
\end{equation}
where $\delta_f$ is the phase of $m_{LR}=|m_{LR}|e^{i \delta_f}$, and
$-\pi/2 \leq \theta_f \leq \pi/2$ is the mixing angle; they are given by
 the relations:
\begin{eqnarray}
\tan 2\theta_f =\frac{2 |m_{LR}| }{m_{LL}^2 -m_{RR}^2 } \qquad ,\qquad 
 \sin\delta_f &=& \frac{\Im (m_{LR}) }{|m_{LR}|}\ \ . \label{mixing}
\end{eqnarray} 
The physical masses, whith $m_{{\tilde{f}}_{1}}< m_{{\tilde{f}}_{2}}$, 
are given by 
\begin{eqnarray}
  m_{{\tilde{f}}_{1,2}}^2 &=& \frac{1}{2}(2 m_f^2 + m_{LL}^2 + m_{RR}^2 
    \mp \sqrt{ (m_{LL}^2 - m_{RR}^2)^2 + 4|m_{LR} m_f|^2 })\ . \label{mass} 
\end{eqnarray}
The interaction of the neutral gauge bosons $\gamma,Z$ and 
the CP-odd Higgs boson
with the sfermion mass eigenstates is described  by the Lagrangian

\begin{eqnarray}
 {\cal L} = & & -i e A^\mu \sum_{i=1,2} Q_f \tilde{f}_i^* \partial_\mu \tilde{f}_i 
-\frac{i g}{c_W} Z^\mu \sum_{i,j=1,2} \{ 
(I_3^f -Q_f s_W^2) R_{j1}^{\tilde{f}} R_{i1}^{\tilde{f}*} -   
Q_f s_W^2 R_{j2}^{\tilde{f}} R_{i2}^{\tilde{f}*} 
\} \tilde{f}_i^* \partial_\mu \tilde{f}_j
\nonumber \\ 
& & 
-\frac{i g}{\sqrt{2}} W^\mu \sum_{i,j=1,2} 
\{ R_{j2}^{\tilde{f}} R_{i2}^{\tilde{f'}*} 
\} \tilde{f'}_i^* \partial_{\mu} \tilde{f}_j
+   \{ g_{A^0\tilde{f}_L\tilde{f}_R^*}
 R_{j2}^{\tilde{f}} R_{i1}^{\tilde{f}*} + 
g_{A^0\tilde{f}_L^*\tilde{f}_R} 
R_{j1}^{\tilde{f}} R_{i2}^{\tilde{f}*}\} A^0 \tilde{f}_i^* \tilde{f}_j
\end{eqnarray}
with
$g_{A^0\tilde{f}_L\tilde{f}_R^*} = 
-\frac{g m_f}{2 m_W} ( A_f (\tan\beta)^{- 2 I_3^f} +\mu^*   )$ and 
$g_{A^0\tilde{f}_L^*\tilde{f}_R} = -(g_{A^0\tilde{f}_L\tilde{f}_R^*})^*$

For the coupling of the CP-odd $A^0$ to a pair of sfermions, one can easily
show that it takes the following form
\begin{eqnarray}
& & A^0 \tilde{f}_1^* \tilde{f}_2=-\frac{g m_f}{2 M_W}
\{ ( A_f (\tan\beta)^{-2 I_3^f} +\mu^* ) e^{-i \delta_f} \sin\theta_f^2 + 
( A_f^* (\tan\beta)^{-2 I_3^f} +\mu ) e^{i \delta_f} \cos\theta_f^2 \}\nonumber\\
& & A^0 \tilde{f}_1^* \tilde{f}_1= \frac{-ig m_f}{2 M_W}\sin 2\theta_f 
\{ |A_f| (\tan\beta)^{-2 I_3^f} \sin ({\rm arg}(A_f) - \delta_f) -
 |\mu| \sin( {\rm arg}(\mu) +\delta_f )\}\nonumber \\
& & A^0 \tilde{f}_2^* \tilde{f}_2  =  - (A^0 \tilde{f}_1^* \tilde{f}_1)\ \ .
\label{A0tt}
\end{eqnarray}
Note that in the MSSM with real soft SUSY breaking parameters, 
the coupling of $A^0$ 
to a pair of sfermions satisfies the following relation:
$(A^0\tilde{f}_i \tilde{f}^*_j)_{i\neq j} = 
-(A^0\tilde{f}_j \tilde{f}^*_i)_{i\neq j}$. The
couplings $(A^0\tilde{f}_i \tilde{f}^*_i)$, i=1,2, vanish
and only $(A^0\tilde{f}_1 \tilde{f}^*_2)= 
- (A^0\tilde{f}_2 \tilde{f}^*_1) $ are non-zero.

\section*{3.~CP-odd production at $e^+e^-$ colliders}
\subsection*{3.1~Associated photon-$A^0$ and Z-$A^0$ production:}

Now we are ready to discuss the associated production mechanisms
$e^+e^- \to A^0 \gamma$ and $e^+e^- \to A^0 Z$ in the MSSM 
with and without CP violating phases. 
These processes have been studied in the MSSM
with real parameters in Refs. \cite{abdel} and \cite{AAC1},
where the full set of Feynman diagrams which 
contribute to these processes can be found.
In both studies it was shown that  
the sfermion contribution to the vertices
$\gamma$-Z-$A^0$ and Z-Z-$A^0$ vanishes due to the fact that 
$(A^0\tilde{f}_1 \tilde{f}^*_2)= -(A^0\tilde{f}_2 \tilde{f}^*_1)$.

In the presence of CP violating phases in $\mu$ and $A_f$,
one can clearly see from (\ref{A0tt}) that $(A^0\tilde{f}_i \tilde{f}^*_i)\ne 0$ for i=1,2, and 
$(A^0\tilde{f}_1 \tilde{f}^*_2)\ne -(A^0\tilde{f}_2 \tilde{f}^*_1)$.
Consequently the sfermion contribution to 
$e^+e^- \to \gamma A^0$ and $e^+e^- \to Z A^0$ (Fig.1) 
no longer vanishes. Note that in the case of stop, the effect of CP 
phases in $(A^0\tilde{t}_i \tilde{t}^*_i)$, i=1,2, 
is enhanced by the top mass as well as by large 
$|A_t|$ and large $|\mu|$. While in the case of scalar 
bottom and tau, the effect of the CP phase of $A_b$ and $A_\tau$
may show up with large $\tan\beta$ eq.(\ref{A0tt}).

\subsection*{3.2~Single $A^0$ production via $e^+e^- \to 
\nu_e\overline \nu_e A^0$}

It is well known that at high energies $\sqrt{s} > 500 $ GeV,
WW fusion  $e^+e^-\to \nu_e\overline \nu_e h^0$ 
is the most promising process with which to discover the Higgs boson.
Due to CP invariance, $A^0$ possesses no
tree-level coupling $A^0$-W-W. The
dominant contribution is therefore from higher order diagrams
which will be mediated by both SM and SUSY particles.
In the MSSM, one can generate  
the $\nu_e\overline \nu_e A^0$ final state at $e^+e^-$ colliders
through one of the generic diagram depicted in Fig.2.
$\nu_e\overline \nu_e A^0$ can also be 
generated through the one-loop process $e^+e^- \to Z A^0$ 
(on-shell Z) followed by the decay of the Z 
boson to $\nu_e\overline \nu_e$, although such a final state could be 
removed experimentally by recoil reconstruction. Consequently,
this two-body production and decay will not be addressed here.

In Fig.2, we show the generic one-loop contributions to 
$e^+e^-\to \nu_e\overline \nu_e A^0$. These comprise the following:
a.) Fig. 2.1: contribution from W-W-$A^0$ vertex 
b.) Fig. 2.2-2.5:  $e^+e^- \to Z^*\to \nu_e\overline \nu_e^*,
\nu_e^*\overline \nu_e $
and $e^+e^- \to W^*\to \nu_e\overline \nu_e^*,\nu_e^*\overline \nu_e $
followed by one-loop neutrino decay: 
$\nu_e^* \to \nu_e A^0$ or 
$\overline\nu_e^* \to \overline\nu_e A^0$
c.) Fig.2.6-2.8 and 2.14 to 2.16: 
one-loop contribution from $e^+e^- \to A^0 Z^*$ with 
$Z^* \to \nu_e \overline \nu_e$, 
d.) Fig. 2.9 and 2.10: 
one-loop correction to $A^0e^+e^*$ and $A^0e^-e^*$,
e.) Fig. 2.11 $\to$ 2.13:
box diagrams with virtual gauge bosons $V=\gamma,Z$
and f.) Fig.2.17$\to$2.22: box diagrams with virtual gauge boson W.
Additional contributions come from 
the non-diagonal self-energies of photon-$A^0$,  
Z-$A^0$ and $W^\pm$-$H^\mp$. However, the amplitudes of these 
contributions are proportional to the electron mass and 
consequently vanish in the approximation $m_e\approx 0$.

All the Feynman diagrams are generated and computed using 
FeynArts \cite{FA} and FormCalc \cite{FC} packages
 in the dimentionnal regularization scheme. We also use 
FF-package and looptools \cite{FF} in the numerical analysis.
In the case of $e^+e^- \to \nu_e \bar{\nu}_e A^0$,
the three body phase space integration is performed using VEGAS 
routines \cite{vega}.

\section*{4. Results}
Before discussing our numerical results, 
we define the free parameters that will be used. 
In MSSM, it is common to parameterize the tree level 
Higgs sector with $\tan\beta$ and $M_A$.
However, in our cases $e^+e^- \to \gamma A^0 \ , \ ZA^0$ and
$e^+e^- \to \nu_e \overline \nu_e A^0$ are one-loop
mediated. In all cases, the amplitudes depend on tree level 
masses and parameters, the inclusion of radiative correction to those 
masses and parameters would be of higher order.
Consequently, in our numerical analysis we will use 
$\tan\beta$ and $M_A$ to parameterize the Higgs sector.
We stress in passing that the cross sections are not very sensitive
to the effect of radiative corrections on the Higgs sector.
We will assume  that $\tan\beta \geq 2.5$.

The chargino neutralino sector can be parametrized by the usual $M_1$, 
$M_2$ and $\mu$. We assume that,  $M_1\approx M_2/2$ and
$\mu > 0$. The third-generation sfermions are parametrized 
by: a common sfermion mass $M_{SUSY}=\widetilde{M}_L=\widetilde{M}_R$,
soft trilinear $A_f$ coupling which we will take to be identical
for top, bottom and tau ($A_t=A_b=A_\tau$), $\tan\beta$ and the 
$\mu$ parameters. Once these parameters are given, 
the mixing angle and the 
sfermions masses are fixed by eqs. (\ref{mixing}, \ref{mass}).
In our analysis we will take into 
account the following constraints when the SUSY parameters are varied:\\
i) the extra contribution $\delta\rho$ to the $\rho$ parameter 
\cite{hagiwara} should not exceed the current limits from 
experimental measurements $\delta\rho \la  10^{-3}$,\\ 
ii) 
$m_{\tilde{t}_1,\tilde{b}_1} > 100$ GeV, $m_{\chi_1^\pm} > 103$ GeV, 
$m_{\chi_1^0} > 50$ GeV and $m_h>110$ GeV.

We first start with the MSSM contributions to 
$e^+e^- \to A^0 \gamma$
and $e^+e^- \to A^0 Z$. We focus only on the case
where the SUSY particles are rather light, taking
$M_{SUSY}=\mu\approx 200$ GeV, $M_2=150$ GeV
and $\tan\beta=2.5$. For such low $\tan\beta$ 
and in order to satisfy $m_h>110$ GeV in MSSM 
with real parameters,
we need large $A_t\approx  6$ TeV and large 
third generation SUSY scale 
$M_{stop}^{L,R}$ of the order 3 TeV together with $M_A>130$ GeV.
While in MSSM with CP phases, the bound $m_h>110$ GeV 
is reduced and even a light Higgs bosons with a mass 
$m_{H_1}\approx 70$ GeV may have escape detection at 
LEPII \cite{MSSMCP1}.
Note that the 2HDM contribution is 
enhanced in the small $\tan\beta \la 1$ regime by 
the top quark mass \cite{AAC1,AAC2,abdel,LG} 
while for large $\tan\beta$ the 2HDM contribution is 
suppressed and does not attain observable rates. 

As can be seen in Fig.3, the 2HDM cross section for both 
$e^+e^- \to A^0 \gamma$ and $e^+e^- \to A^0 Z$ is enhanced by 
light SUSY particles. The SUSY enhancement is not that 
spectacular since there is strong destructive interference
between the vertex and box diagrams which reduces the cross section.
In the case of $e^+e^- \to A^0 \gamma$ one can reach 
$\approx 0.02$ fb for $M_A<350$ GeV. 
The cross section can also be enhanced by polarizing the electron 
and positron beams. As shown in Fig.3,
the cross section with left-handed longitudinally
polarized electrons is approximately twice 
that for the non-polarized case, while for 
left-handed electrons and right-handed positrons 
the enhancement is approximately a factor of 4. The cross
section of $e^+e^- \to A^0 Z$ is approximately
one order of magnitude smaller than $e^+e^- \to A^0 \gamma$, and
consequently $e^+e^- \to A^0 \gamma$ is more 
promising, especially if polarized electron positron 
beams are available. 
At a higher center of mass energy ($\sqrt{s}=800$ GeV) with light SUSY
particles $M_{SUSY}=200$ GeV,
the cross sections for $e^+e^- \to A^0 \gamma, A^0 Z$ are suppressed. 
The maximum value for $e^+e^- \to A^0 \gamma$ (resp $e^+e^- \to A^0 Z$)
is about 0.003 fb (resp 0.001 fb) for $300 <M_A <350$ GeV.

Let us now discuss the single CP-odd Higgs boson production
$e^+e^- \to \nu_e \bar{\nu}_e A^0$. 
We start with the 2HDM contribution
which basically arises from the typical diagrams shown in 
Fig.2.1, 2.6, 2.11$\to$ 2.20.
The dominant contribution comes from top-bottom contribution 
in Fig.2.1 and 2.6 with fermion exchange
\footnote{It has been shown recently \cite{Eberl}, that the 
radiative correction to $e^+e^- \to \nu_e \bar{\nu}_e h^0$ in 
the MSSM receive sizeable enhancements from heavy fermion loops 
compared to sfermion loops.}.
Ref. \cite{LG} evaluated the top-bottom contribution to 
$e^+e^- \to \nu_e \bar{\nu}_e A^0$ coming from Fig.2.1 only. 
The full 2HDM contribution is under investigation, 
and preliminary results can be found in \cite{SU}.
We have cross checked with the existing results \cite{LG,SU}
and we found agreement.

In Fig.~4, we plot the cross section of $e^+e^- \to \nu_e \bar{\nu}_e A^0$ in the 2HDM for 500 GeV and 800 GeV center of mass energies. 
For the Higgs couplings and masses we use the MSSM 
values including radiative corrections to the lightest Higgs boson mass.
We found that the inclusion of the diagrams of Fig.2.6, which was neglected in \cite{LG}, can slightly enhance
the cross section which can reach 0.0033 fb at $\sqrt{s}=500$ GeV,
$M_A=100$ GeV  and small $\tan\beta=0.5$. 
For large $\tan\beta$ the cross section drops by 2 
order of magnitude. At high energy $\sqrt{s}=800$ GeV,
the cross sections are of the order 0.006 fb for small 
$\tan\beta=0.5$ and  $M_A<400$ GeV.\\
We have also included the full MSSM contribution 
coming from  vertex and boxes (Fig.2),
omitting the five point-functions, these
five point-functions do not have any enhancement factor and 
their contribution is expected to be smaller.\\
Numerically, in the MSSM with real parameters, we found that even 
in the optimistic scenario where all
SUSY particles are light of order 200 GeV and $\tan\beta=2.5$
the cross section does not receive any substantial enhancement
both at $\sqrt{s}=500$ GeV and $800$ GeV.

Let us now turn on the CP violating phases of the trilinear terms 
$A_t$, $A_b$ and $A_\tau$ and see their effect on 
$e^+e^- \to \gamma A^0$, $e^+e^- \to Z A^0$ and $e^+e^- \to \nu_e 
\overline{\nu}_e A^0$ .
As was pointed in the introduction, allowing large CP violating 
phases in $A_t$ and $A_b$ may violate the experimental bounds on the 
electron and neutron EMDs \cite{2-loop}. Since we are interested 
only in the effect of the CP phases of third generation 
sfermions\footnote{The coupling of $A^0$ to sfermions is 
proportional the fermion mass. Consequently
the first and second generation do not have any impact 
on the diagrams of Fig.~1 with sfermion exchange.}, 
the first and second generation sfermions can be decoupled 
together by taking them at 5 TeV. 
Then the one-loop electron and neutron EDMs \cite{savoy} 
are safely within the experimental limits.
However, in the scenario of large CP phases in 
$A_t$, $A_b$ and $A_\tau$,
two-loop Barr-Zee type diagrams can violate the EDMs 
constraints for large $\tan\beta\ga 30$. In such a case one may 
need to arrange for a cancellation mechanism among the various 
one and two-loop contributions or among the full two-loop contribution
alone. In our study we limit ourself to the case 
of moderate $\tan\beta \leq 25$.

It has been shown in \cite{MSSMCP0}, that when the three neutral Higgs 
bosons mix their effective couplings to fermions can be 
rather different at one loop. 
However, one can find parameters space in the MSSM where 
such corrections turn out to be only of the 
order of few percent \cite{BR0}, in such case one of 
the $H_i^0$ look like CP-odd $A^0$.
Taking into account the above observation,
we present our numerical results for 
$e^+e^- \to \gamma A^0\ , \ Z A^0$ and
$e^+e^- \to \nu_e \overline \nu_e A^0$
only in the case of low $|\mu| \leq 600$ GeV, 
low $|A_{t,b}|\leq 600$ GeV and low SUSY scale 
$M_{SUSY}\leq 300$ GeV where scalar-pseudoscalar mixing is not
sizeable and so such approximation is correct. 
In our numerical results
we assume that only $A_{t,b,\tau}$ carry CP violating phases.
We assume that $\mu$, $M_1$ and $M_2$ are real, since we do 
not expect the charginos neutralinos contributions
to give sizeable enhancement.\\
As we stress above, the first and second generation 
sleptons and squarks are set to 5 TeV. We have observed that there 
is some non-decoupling effects in $e^+e^- \to A^0 \gamma , A^0 Z$ and 
$e^+e^- \to \nu_e \overline{\nu}_e A^0$. 
Those non-decoupling effects originate from 
t--channel diagrams with exchange of two neutralinos(charginos) 
and one selectrons(sneutrinos). The asymptotic dependence on sleptons 
mass $m_{\tilde l}$ goes like ${\rm log}(m_{\tilde l})$
for very large $m_{\tilde l}$ compared to charginos and neutralinos
masses. This non-decoupling effect can be seen in Fig.~5 
only for low $\mu=300$ GeV ad gives 0.025 fb (resp 0.015 fb) 
cross section for  $e^+e^- \to A^0 \gamma$ (resp $e^+e^- \to A^0 Z$) at 
small CP phases.
\\
Numerically, it turns out that the squark contribution with 
virtual photon exchange (Fig.1) is more important than
the one with virtual Z exchange for both $e^+e^- \to A^0 \gamma$ and 
$e^+e^- \to A^0 Z$. Note that the couplings $Z\tilde{t}_i\tilde{t}_j^*$
are not very sensitive to the CP phases while $A^0\tilde{t}_i\tilde{t}_j^*$
exhibit a strong dependence on the CP phases. For instance, the coupling 
$A^0\tilde{t}_i\tilde{t}_i^*$, which is zero for vanishing CP phases,
reaches its maximum value, which is comparable to 
$A^0\tilde{t}_1\tilde{t}_2^*$, for arg($A_t) \ga 5$ degrees.

We illustrate in Fig.~5 the cross section of $e^+e^- \to A^0 \gamma$
(left plot) and $e^+e^- \to A^0 Z$ (right plot) as function of CP phases
 ${\rm arg}(A_f)={\rm arg}(A_t)={\rm arg}(A_b)={\rm arg}(A_\tau)$ at 
$\sqrt{s}=500$ GeV, $\tan\beta=2.5$ and for $\mu=300,450, 600$  GeV. 
The CP phase is varied from 0 to about 100 degrees. 
For CP phases greater that $100$ degrees, the mass of the 
lightest stop $m_{\tilde{t}_1}$  
becomes less than about $100$ GeV which is ruled out by experiments.
As one can see in Fig.~5, for low value of $\mu$ the effect of 
the CP phase ${\rm arg}(A_f)$ is not important.\\
In both cases, the cross section can reach values of about 0.08 fb
for large CP phases and $\mu=600$ GeV. The sensitivity to the 
CP phase is more important for $e^+e^- \to A^0 Z$. 
This is simply due to the fact that gauge 
invariance in case of $e^+e^- \to A^0 \gamma$ allows the photon 
to couple to only two identical sfermions with the 
electromagnetic coupling. 
For large $\tan\beta$, the contributions of the CP violating phases of 
$A_b$ and $A_\tau$ may show up and becomes comparable to the effect of 
the CP phase of $A_t$. 
We have checked that for $\tan\beta\approx 24$, 
$\sqrt{s}=500$ GeV, light CP-odd $M_A=200$ GeV and 
large CP phases arg$(A_f)\approx \pi/2$,
the maximum reach for $e^+e^- \to A^0 \gamma$ is about 
$0.032$ fb, decreasing for larger $M_A$. 
In the case of $e^+e^- \to A^0 Z$ production,
the situation is better. A cross section of about 0.052 fb 
can be obtained for $\tan\beta=24$, 
large CP phases arg$(A_f)\approx\pi/2$ and $M_A=200$ GeV. At large
$\tan\beta$, the contribution from the CP phases of 
$A_\tau$ and $A_b$ can give cross sections comparable to
those from the CP phase of $A_t$ 
in the low $\tan\beta$ regime, especially in the case of 
$e^+e^- \to A^0 Z$ production.\\
In Fig.~6, we show the CP phases effect on $e^+e^- 
\to \nu_e \overline{\nu}_e A^0$ for center of mass energy $\sqrt{s}=500$ GeV
 , $M_A=200$ GeV (left) and $\sqrt{s}=800$ GeV, $M_A=400$ GeV (right).
In both cases, the large CP phases can enhance the cross section by
one to two order of magnitude and one can reach 0.08 fb cross sections
for 500 GeV center of mass energy.

To end this analysis, for completeness, we 
present cross sections for $e^+e^- \to ZH_i^0$ and 
$e^+e^- \to \nu_e \bar{\nu}_e H_i^0$
in CP violating MSSM \cite{MSSMCP2,MSSMCP1}. 
We use the same set of parameters used in $e^+e^- \to ZA^0$ and
$e^+e^- \to \nu_e \bar{\nu}_e A^0$:
\begin{eqnarray}
M_{SUSY}=\widetilde{M}_L=\widetilde{M}_R=M_2=300\ , \ \mu =300 \to 600 \ , \ 
|A_t|=|A_b|=600\ \ GeV\label{para}
\end{eqnarray}
In CP violating MSSM we use the public fortran program \cite{cph}.
We will take the charged Higgs mass $M_{H^\pm}=215$ GeV as input parameter. For this charged Higgs mass and the 
set of parameters fixed above,
$H_2^0$ is  dominated by CP-odd component and it's mass  
 about 200 GeV for small CP phases while $H_3$ is dominated 
by CP-even component and it's mass about 220 GeV.

In Fig.6 we show cross section for 
$e^+e^- \to ZH_i^0$ at $\sqrt{s}=500$ GeV, 
$\tan\beta=2.5$ (left plot) and WW fusion
$e^+e^- \to \nu_e \bar{\nu}_e H_i^0$ at high energy 
$\sqrt{s}=800$ GeV (right plot).
It can be seen from the left  plot that for 
$\mu=300$ GeV (solid lines), the cross section of 
$e^+e^- \to ZH_2^0$ (resp $e^+e^- \to ZH_3^0$) 
is less than $2 \times 10^{-2}$ fb (resp  about 1.2 fb) for 
a large CP phases range. This means that $H_2^0$
is dominated by CP-odd component and $H_3^0$ is dominated 
by CP-even component. For $\mu=600$ GeV the cross section 
of $e^+e^- \to ZH_2^0$ is enhanced  and reach 0.073 fb.
The cross section we obtain in this scenario,
with scalar-pseudoscalar mixing, both for 
$\mu=300$ GeV and $\mu=600$ GeV
are comparable to what we found for 
$e^+e^- \to Z A^0$ with large CP phases.
At $\sqrt{s}=800$ GeV, the WW fusion offers better reach for
$e^+e^- \to \nu_e \bar{\nu}_e H_2^0$ with 0.2 fb cross section.

\section*{5. Conclusions}
We have computed the cross--sections for the production
mechanisms $e^+e^-\to A^0\gamma$, $e^+e^-\to A^0Z$
and  $e^+e^- \to A^0  \nu_e\bar{\nu}_e$ at high energy $e^+e^-$ 
colliders in the framework of the MSSM. 
Such processes proceed via higher order
diagrams and are strongly model dependent. We presented
results for the 2HDM and the MSSM. 
\\
In the 2HDM the cross sections are small for large $\tan\beta$
and are enhanced by the top loop effect for $0.5 < \tan\beta <1$.
In the MSSM with real parameters, light SUSY particles may 
enhance the cross sections resulting in maximum values 
of order 0.02 fb for $e^+e^- \to A^0 \gamma$, which can be 
enhanced by a factor of 2 if the electron beam can be polarized.

In the case of $e^+e^-\to A^0Z$
and  $e^+e^- \to A^0  \nu_e\bar{\nu}_e$,
light SUSY particles may give important contributions
to the cross--section, resulting in maximum values 
of order 0.005 fb (resp 0.001 fb) for $e^+e^-\to A^0Z$ (resp 
$e^+e^- \to A^0  \nu_e\bar{\nu}_e$) for 
$\tan\beta=2.5$ and $\sqrt{s}=500$ GeV. 
Therefore the SUSY enhancement is not sufficient to produce
an observable signal at the planned luminosities 
of $500 {\mbox{fb}}^{-1}$. 
With an integrated luminosity of 
${\cal L}\approx$ 500-1000 ${\mbox{fb}}^{-1}$ expected at
proposed $e^+e^-$ colliders, in the MSSM with real parameters 
one could expect a non--negligible number of events only from 
$e^+e^-\to A^0 \gamma$. While  $e^+e^-\to A^0Z$
and  $e^+e^- \to A^0  \nu_e\bar{\nu}_e$ does not attain
observable rates.\\

In the MSSM with explicit CP phases and small 
scalar-pseudoscalar mixing, one of $H_i^0$ would be dominated by 
CP-odd $A^0$ component,
we have shown that, in this case, $e^+e^- \to \gamma A^0\ , \ ZA^0$ 
and $e^+e^- \to \nu_e \overline{\nu}_e A^0$ processes 
discussed above can be enhanced by large CP phases
 and reach cross sections of the order $0.08$ fb at high energy.
In the MSSM with sizeable scalar-pseudo scalar mixing, 
it is possible to have cross sections greater than $0.1$ fb 
for $e^+e^- \to Z H_i$ and 
$e^+e^- \to \nu_e \bar{\nu}_e H_i$ for all i=1,2,3. \cite{AA}.
Therefore such signals could not be explained in the MSSM unless 
 sizeable scalar-pseudoscalar mixing is taken place.

\acknowledgements

We are grateful to Thomas Hahn and Christian Schappacher for their 
patience in answering all my questions about the new version of 
FeynArts \cite{FC}. We are grateful to Andrew Akeroyd and to 
Wolfgang Hollik for fruitful discussions and for reading the paper.
We would like to thank Sven Heinemeyer for fruitful discussion.
This work is supported  by Alexander von Humboldt Fondation.

\begin{figure}[t!]
\smallskip\smallskip 
\centerline{{
\epsfxsize4.8 in 
\epsffile{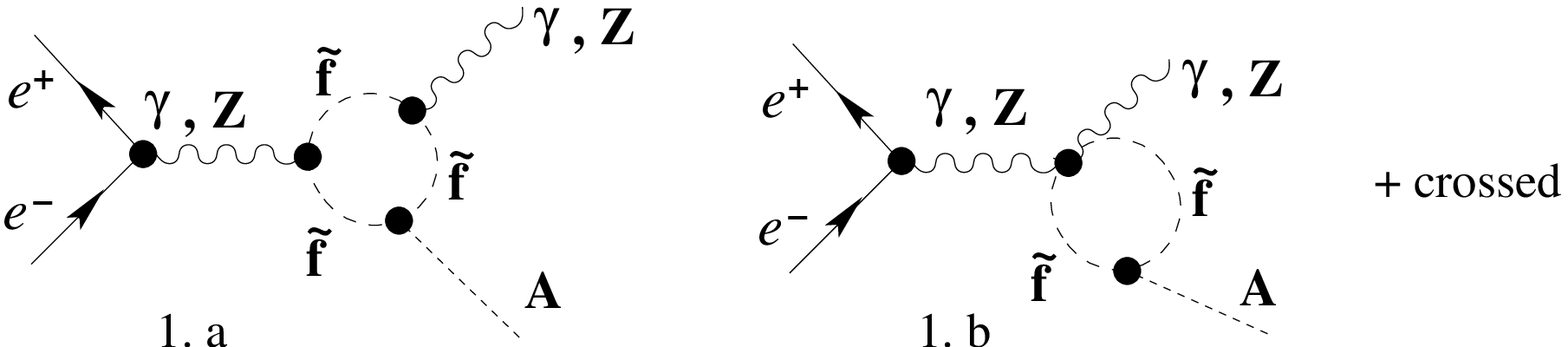}}}
\smallskip\smallskip
\caption{Squark contributions to $e^+e^- \to \gamma A^0$.}
\label{fig1}
\end{figure}
\begin{figure}[t!]
\smallskip\smallskip 
\centerline{{
\epsfxsize6.1 in 
\epsffile{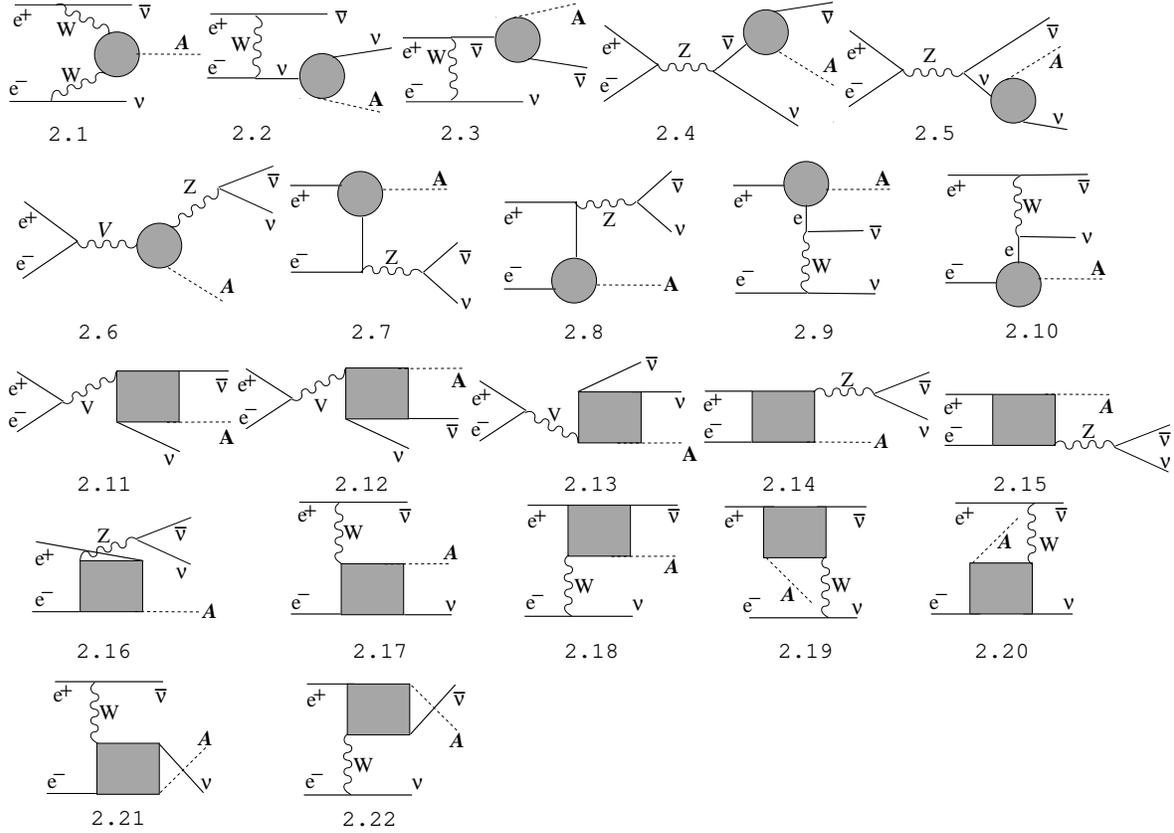}}}
\smallskip\smallskip
\caption{Typical Feynman diagrams contributing to 
$e^+e^-\to \nu \bar{\nu} A^0$ in the MSSM.}
\label{fey2}
\end{figure}
\begin{figure}[t!]
\smallskip\smallskip 
\vskip-5.cm
\centerline{{
\epsfxsize3.1 in 
\epsffile{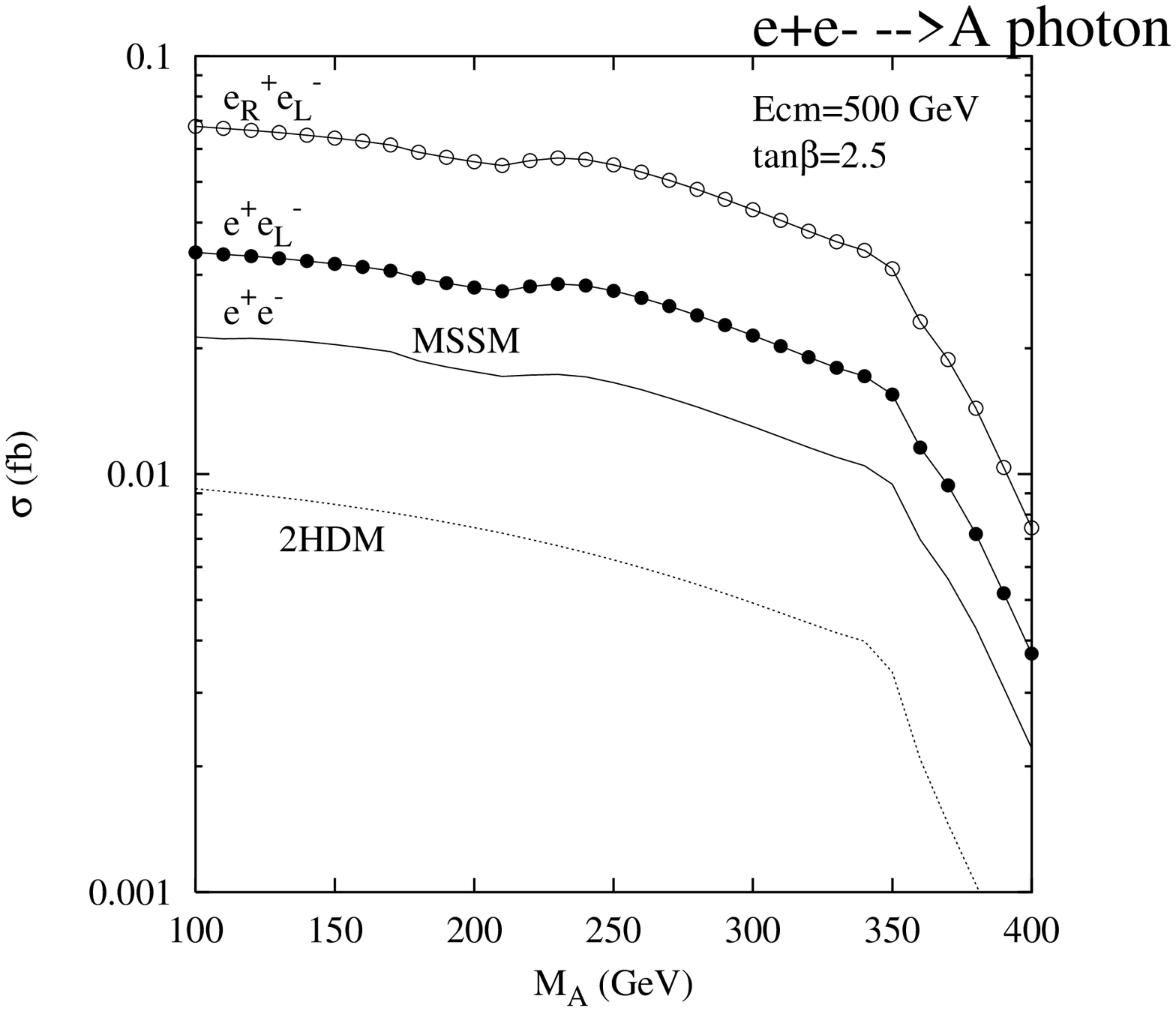}}  \hskip0.4cm
\epsfxsize3.1 in 
\epsffile{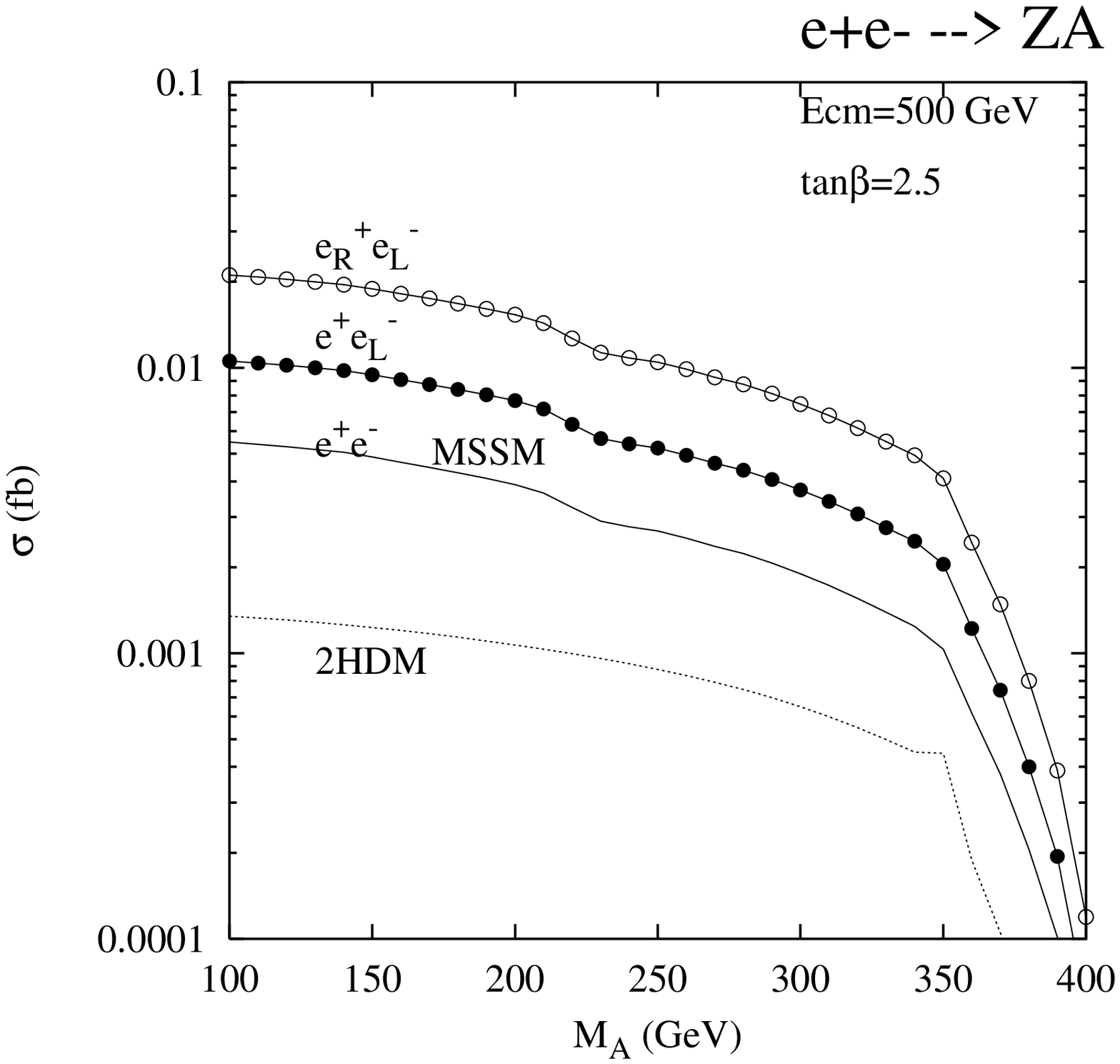} }
\smallskip\smallskip
\caption{Total cross section at 500 GeV center--of--mass energy 
for $e^+e^- \to A^0  \gamma$ (left) and $e^+e^- \to A^0 Z$ (right)
in 2HDM and MSSM with real parameters. 
$M_{SUSY}=M_{\rm sfermions}=\mu\approx 200$ GeV, $M_2=150$ GeV
and $\tan\beta=2.5$}
\label{fey3}
\end{figure}
\begin{figure}[t!]
\smallskip\smallskip 
\vspace{-5.cm}
\centerline{{
\epsfxsize3.1 in 
\epsffile{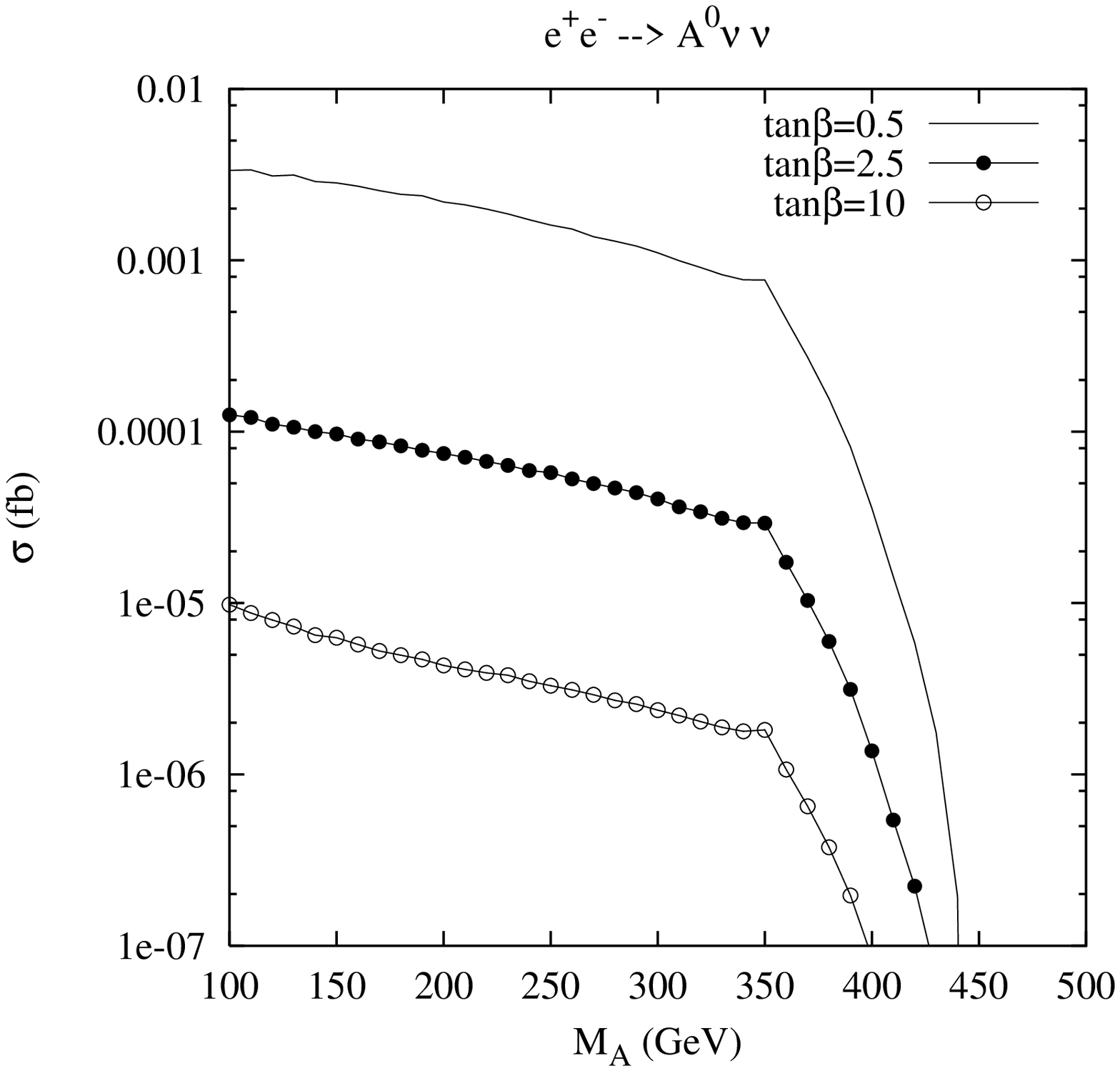}}  \hskip0.4cm
\epsfxsize3.1 in 
\epsffile{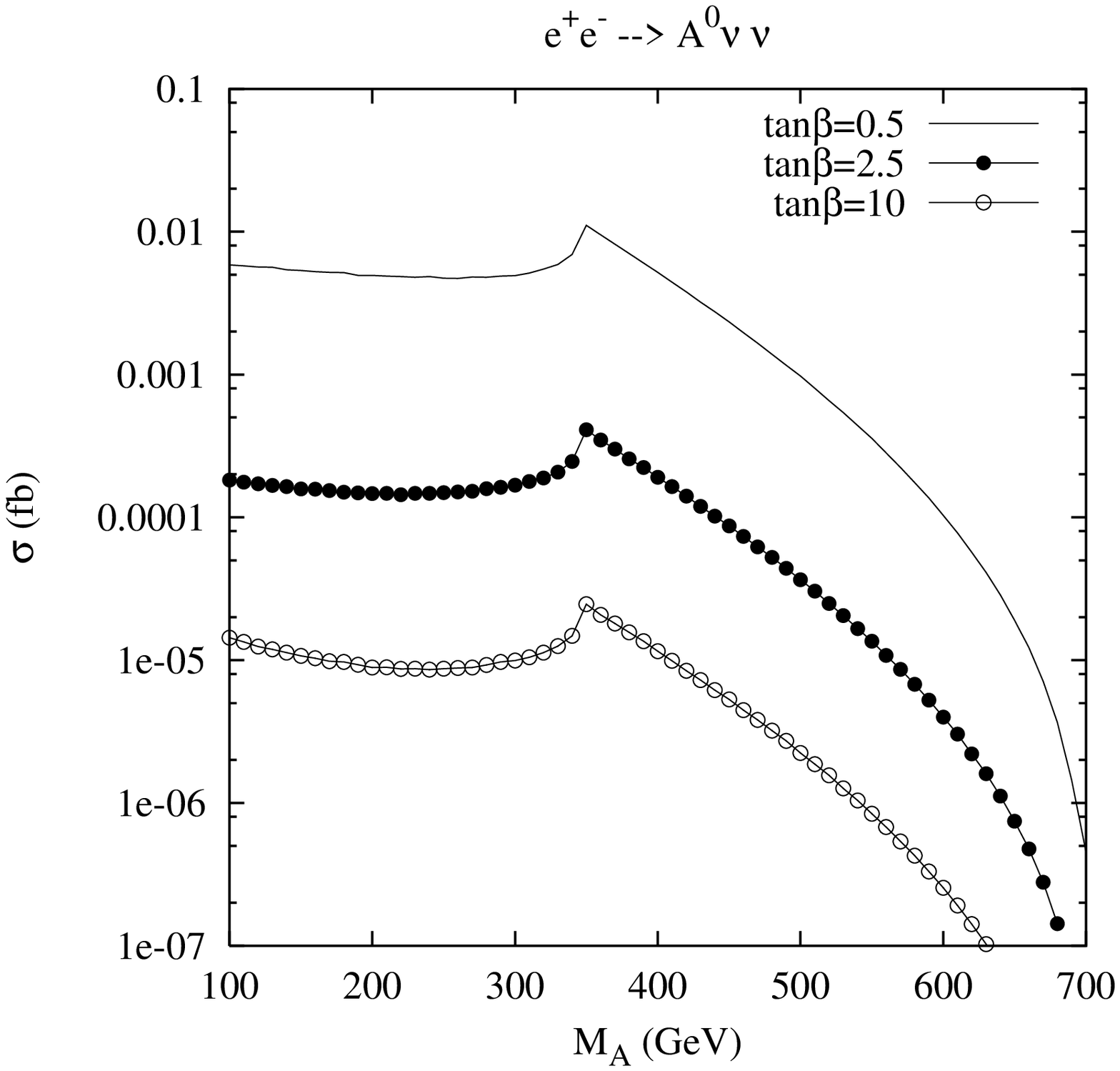} }
\smallskip\smallskip
\caption{Total cross section for $e^+e^- \to A^0  \nu\bar{\nu}$ at 
500 GeV (left) and 800 GeV (right) center--of--mass energy in the 2HDM.}
\label{fey5}
\end{figure}
\begin{figure}[t!]
\smallskip\smallskip 
\vspace{-5.cm}
\centerline{{
\epsfxsize3.1 in 
\epsffile{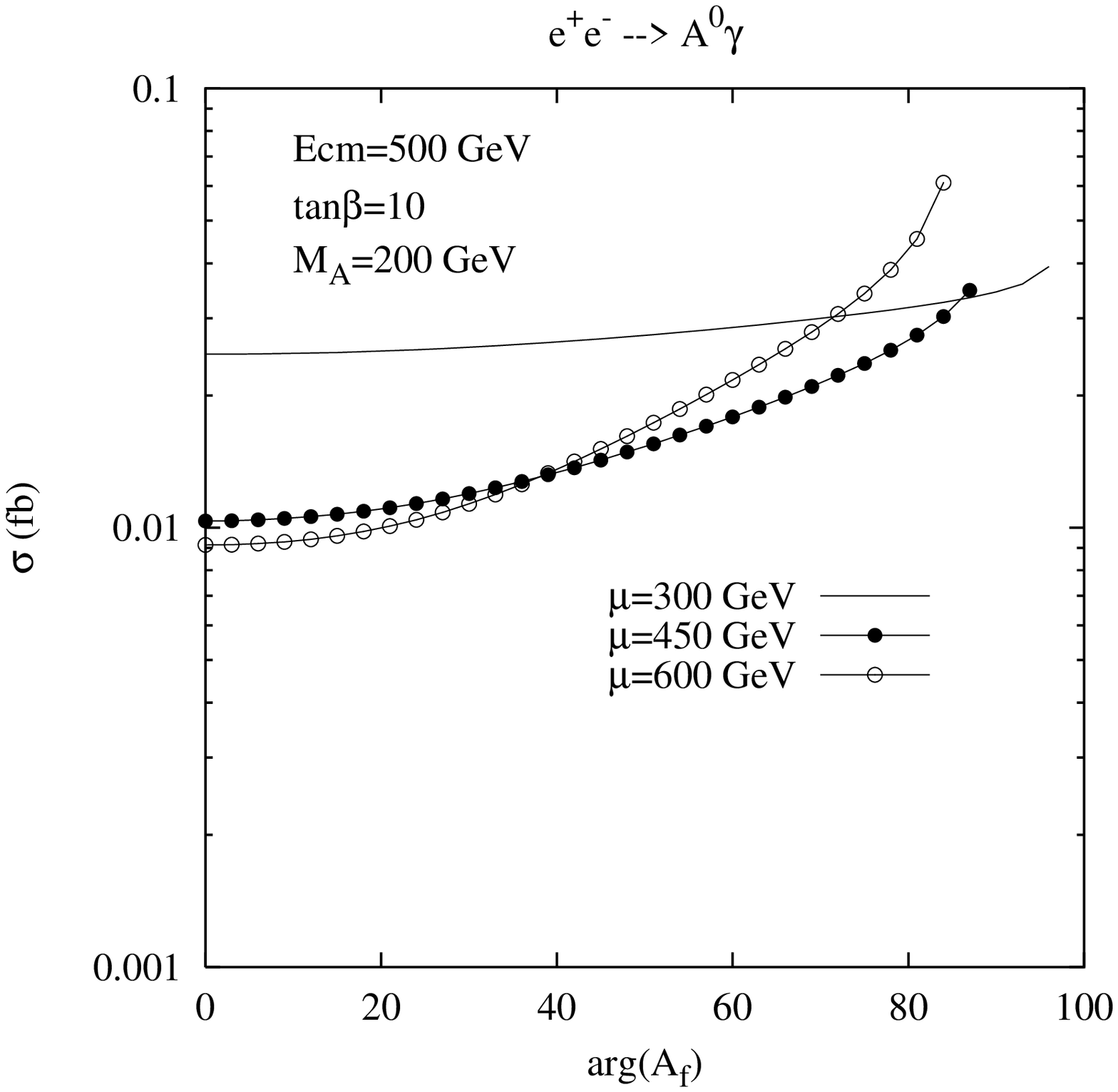}}  \hskip0.4cm
\epsfxsize3.1 in 
\epsffile{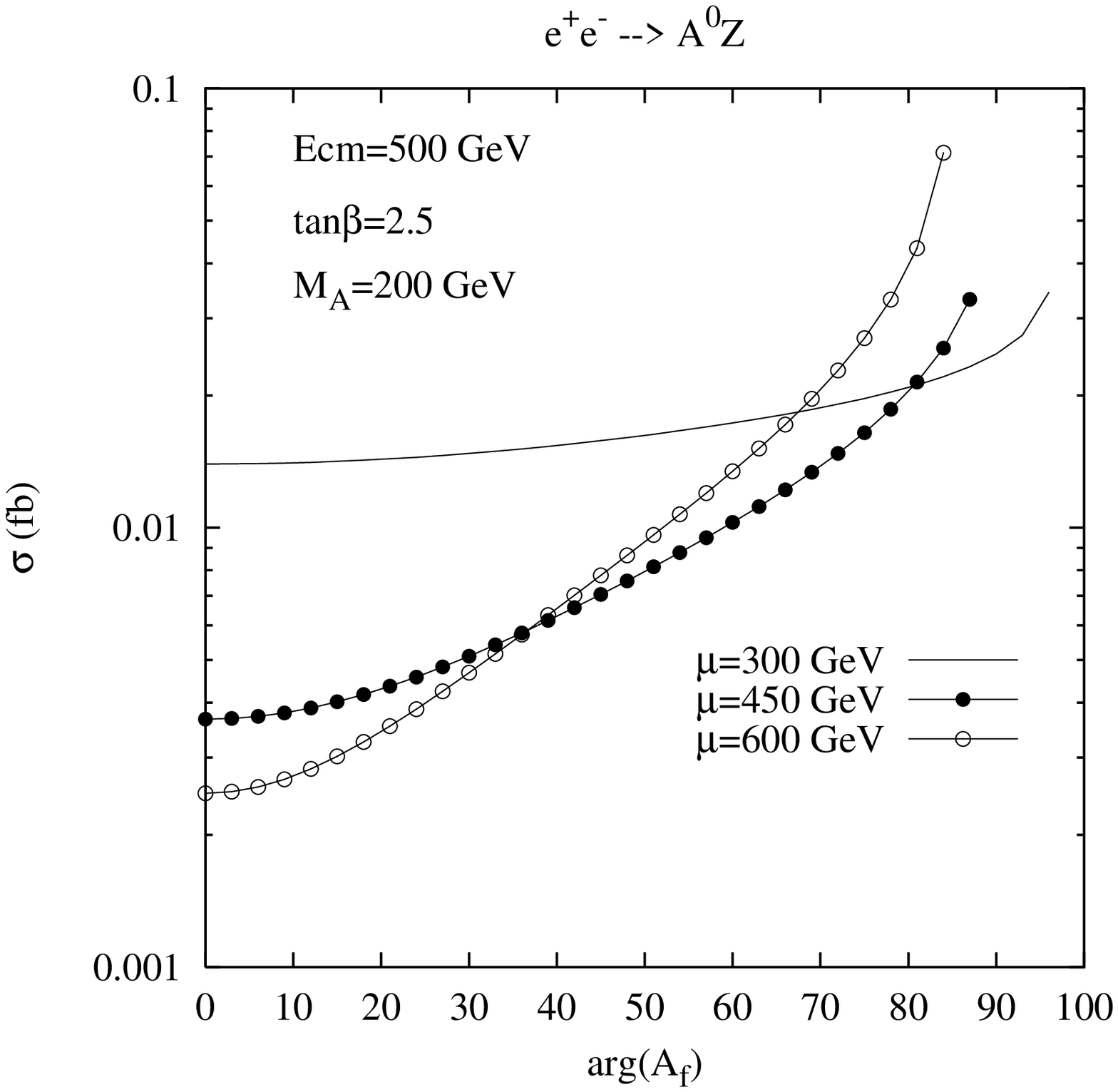} }
\smallskip\smallskip
\caption{Total cross section at 500 GeV center--of--mass energy for 
$e^+e^- \to A^0  \gamma$ (left) and $e^+e^- \to A^0 Z$ (right) 
as function of ${\rm Arg}(A_f)$ for 
$\tan\beta=2.5$  with $M_{SUSY}=300$, 
$A_{t,b}=600$ GeV and $M_2=300$ GeV.}
\label{fey4}
\end{figure}
\begin{figure}[t!]
\smallskip\smallskip 
\vspace{-5.cm}
\centerline{{
\epsfxsize3.1 in 
\epsffile{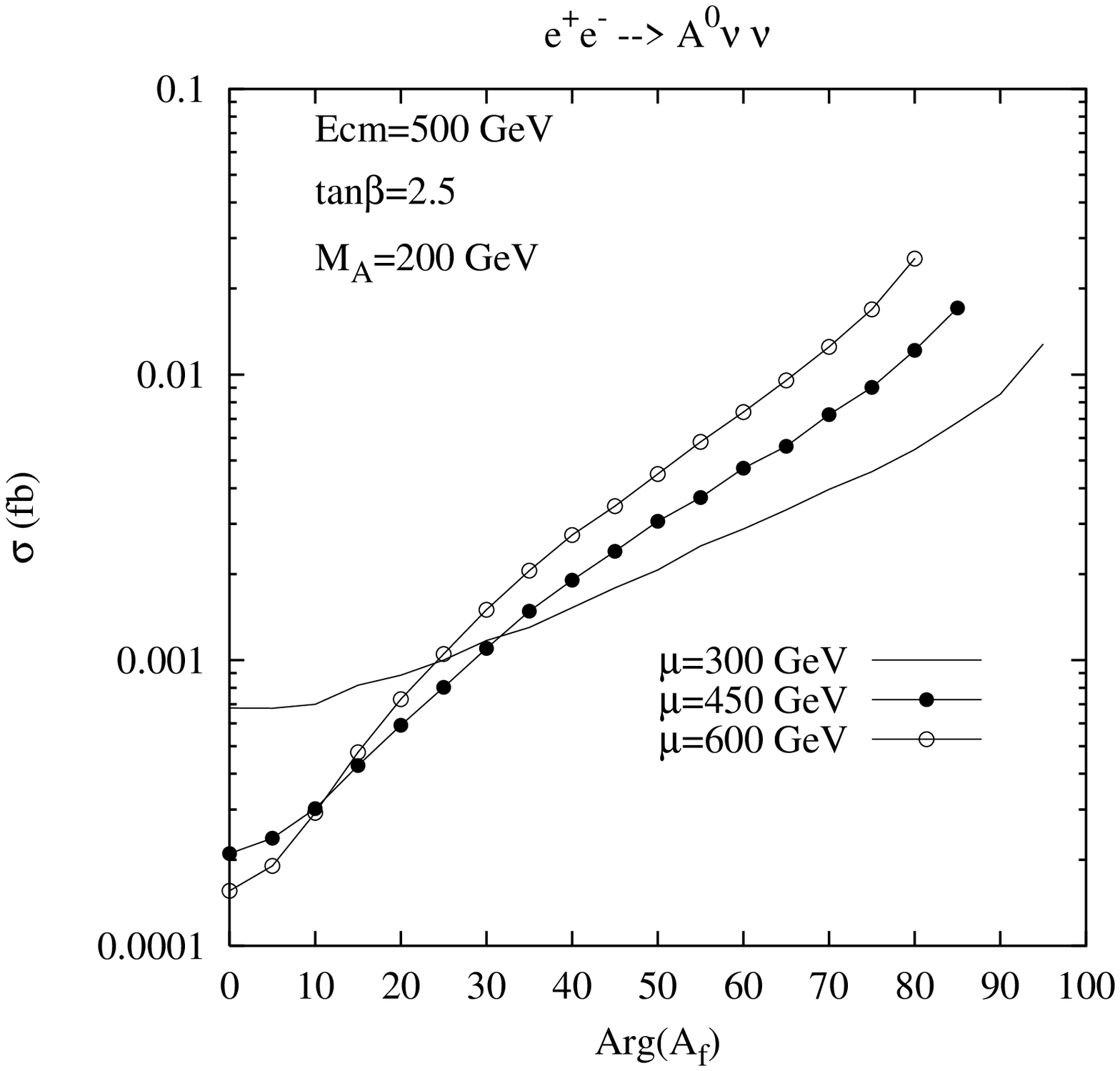}}  \hskip0.4cm
\epsfxsize3.1 in 
\epsffile{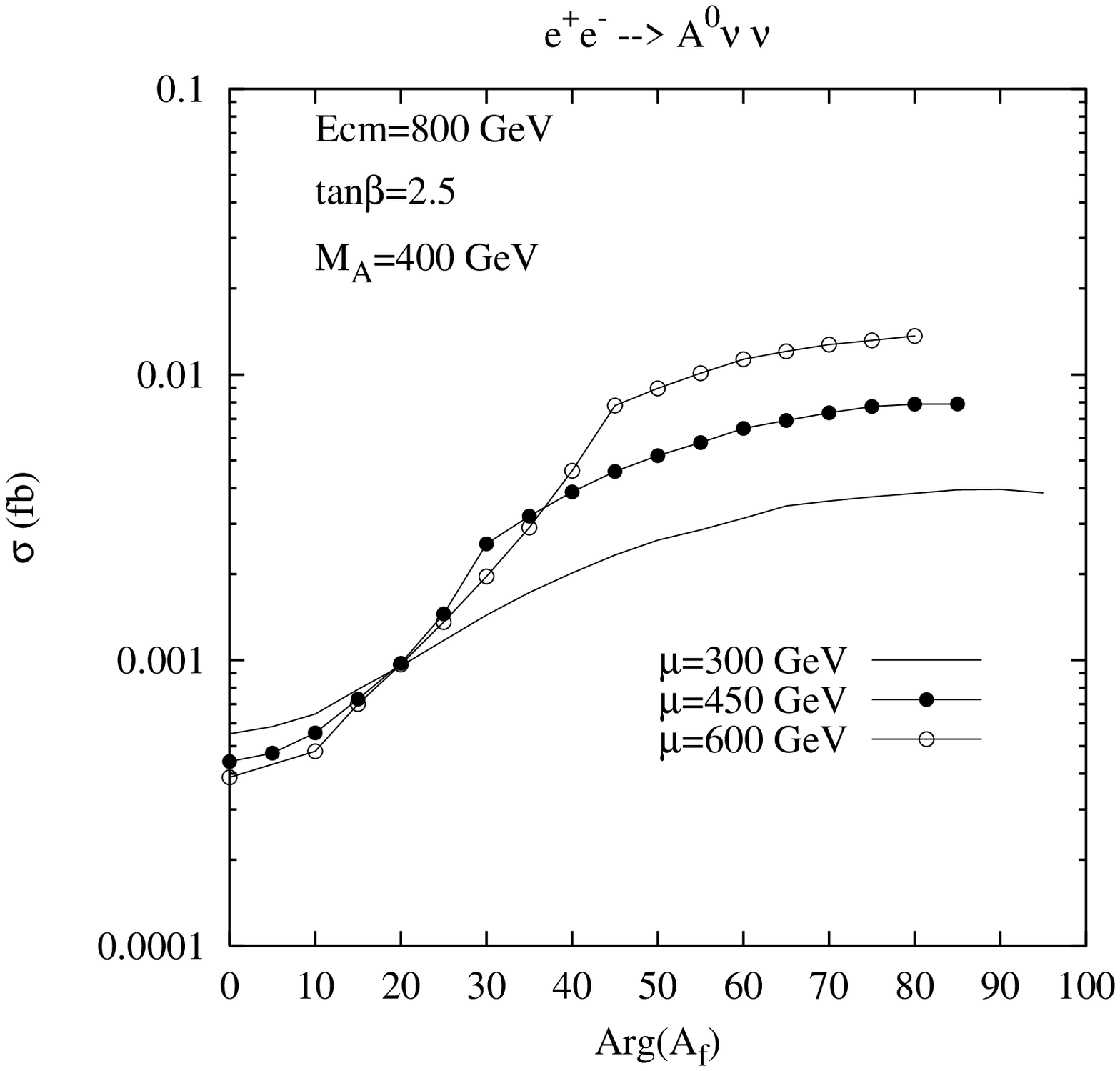} }
\smallskip\smallskip
\caption{Total cross section  for 
$e^+e^- \to \nu_e \overline{\nu}_e A^0$ 
as function of ${\rm Arg}(A_f)$ for 
$\tan\beta=2.5$, $M_{SUSY}=300$, 
$|A_{t,b,\tau}|=600$ GeV and $M_2=300$ GeV. Left plot $\sqrt{s}=500$ GeV and 
right plot $\sqrt{s}=800$ GeV }
\label{fey4}
\end{figure}
\begin{figure}[t!]
\smallskip\smallskip 
\vspace{-5.cm}
\centerline{{
\epsfxsize3.1 in 
\epsffile{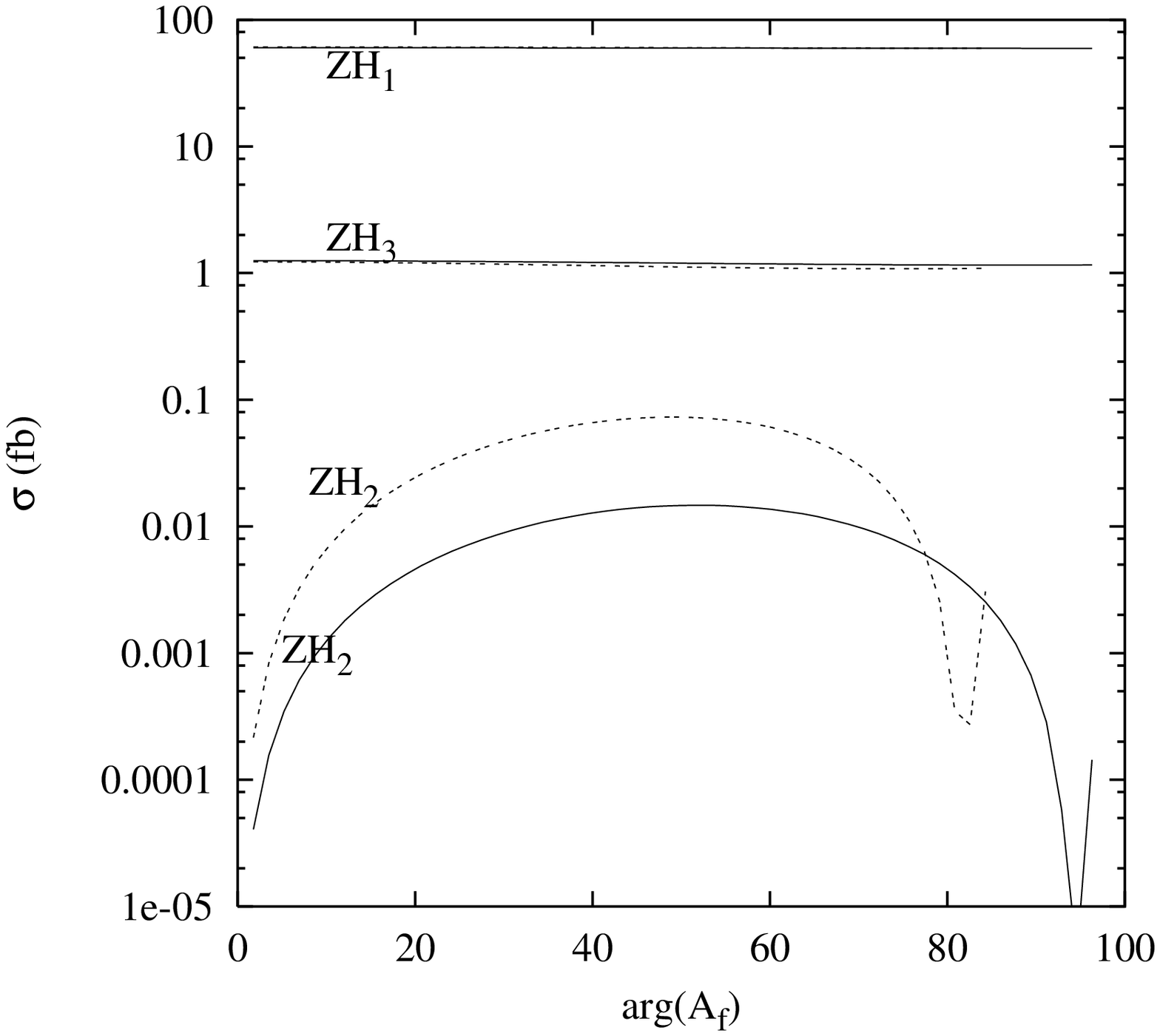}}  \hskip0.4cm
\epsfxsize3.1 in 
\epsffile{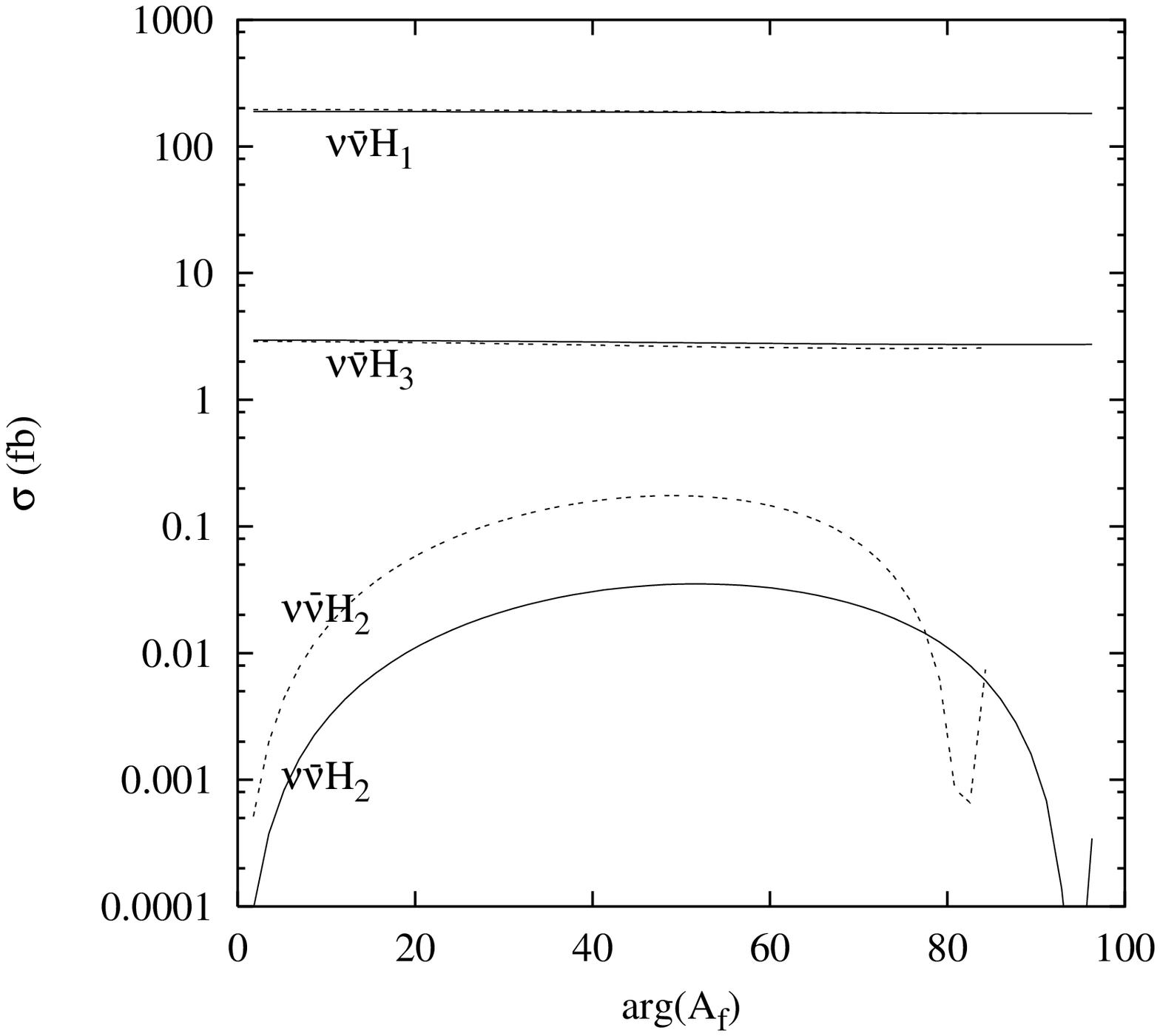}}
\smallskip\smallskip
\caption{Total cross section for: 
(left plot) $e^+e^- \to ZH_i$ at  $\sqrt{s}=500$ GeV,
(right plot) $e^+e^- \to \nu\bar{\nu} H_i$ at  
$\sqrt{s}=800$ GeV with $M_{H^\pm}=215$ GeV and $\tan\beta=2.5$,
$\mu=300$ GeV (solid lines) and $\mu=600$ GeV (dashed lines).
 In both cases we use parameters set in  (\ref{para}) }
\label{fey7}
\end{figure}
\end{document}